\documentclass[
prd,
superscriptaddress,
10 pt,
twocolumn,
preprintnumbers,
notitlepage,
secnumarabic,
nobibnotes,
nofootinbib,
showpacs
]{revtex4-1}

\usepackage[T1]{fontenc}
\usepackage[pdftex]{color,graphicx}
\usepackage{mathtools}
\usepackage{amssymb}
\usepackage{bm}
\usepackage{dsfont}
\usepackage{mathrsfs}
\usepackage{calligra}
\usepackage{tabularx}
\usepackage[artemisia]{textgreek}
\usepackage{hyperref}
\usepackage{xcolor}
\usepackage{enumerate}
\usepackage{multirow}
\usepackage{url}
\usepackage[normalem]{ulem}

\usepackage{perpage}
\MakePerPage{footnote}
\renewcommand{\.}{\!\;}		
\renewcommand{\@}{\!\:\!}	

\DeclareMathAlphabet\mathbfcal{OMS}{cmsy}{b}{n}

\renewcommand{\i}{\!\:\mathrm{i}}

\newcommand{\enn}{\mathcall{\Large{n}\!\.}}
\newcommand{\mathcall}[1]{\text{\calligra\footnotesize #1\,}}

\newcommand{\pos}[1]{\ensuremath{\@\langle#1\rangle}}

\usepackage{tikz}
\usetikzlibrary{shapes.geometric}
\tikzset{cb15/.style={draw=black, line width=1.5pt}}
\tikzset{cb25/.style={draw=black, line width=2.5pt}}

\begin{document}
\title{Continuously distributed holonomy-flux algebra}

\author{Jakub Bilski}
\email{bilski@zjut.edu.cn}
\affiliation{Institute for Theoretical Physics and Cosmology, Zhejiang University of Technology, 310023 Hangzhou, China}


\begin{abstract}
\noindent
The procedure of the holonomy-flux algebra construction along a piecewise linear path, which consists of a countably infinite number of pieces, is described in this article. The related construction approximates the continuous distribution of the holonomy-flux algebra location along a smooth link arbitrarily well. The presented method requires the densitized dreibein flux and the corresponding operator redefinition. The derived result allows to formulate the gravitational Hamiltonian constraint regularization by applying the Thiemann technique adjusted to a piecewise linear lattice. By using the improved Ashtekar connection holonomy representation, which is more accurate than the one used in canonical loop quantum gravity, the corrections related to the redefined densitized dreibein flux vanish. In this latter case, the Poisson brackets of the continuously distributed holonomy-flux algebra along a link between a pair of nodes are equal to the brackets for these smeared variables located at the nodes.
\end{abstract}

\maketitle


\section{Introduction}\label{Sec_Introduction}

\noindent
Loop quantum gravity (LQG) \cite{Thiemann:1996aw,Thiemann:2007zz} is a lattice theory expressed in terms of honolomies and fluxes of the Ashtekar variables \cite{Ashtekar:1986yd}. The considered lattice is constructed as a finite collection of links and nodes. The former objects are the oriented graph's edges with attached SU$(2)$ holonomies. They intersect at the nodes defined as the graph's vertices with attached intertwiners. These latter objects implement the SU$(2)$ invariance on the lattice by averaging possibly different group elements of adjacent (intersecting) links \cite{Thiemann:1996hw,Thiemann:1997rv}. It is believed that in the large links number limit this discrete system approximates the standard continuous formulation of the Einstein gravitational theory \cite{Einstein:1916vd}. This limit is controlled by the small values of the parameters that determine the lengths of the links (or by the parameter if one assumes an equal length). All the aforementioned structure specifies how the graph is embedded as the lattice into the spatial manifold, the geometry of which it represents.

The action of the flux operator in LQG modifies the SU$(2)$ elements at the points where the surface on which this operator is defined intersects with links. To preserve the gauge invariance along the links, these points have to be identified as the nodes with appropriate intertwiners attached. It is worth noting that this feature of the model prevents the values of regulators in the Hamiltonian constraint from being taken to zero. These values determine the two points splitting; hence they are associated with the lengths of the links, for instance, by identifying these lengths with the regulators. If it would be possible to send the regulators to zero, the intersection points of the links and surfaces would need to be densely-defined on the lattice. This case would lead to a pathological structure. Therefore, the constructional restriction that permits taking the value of regulator to zero is introduced by the aforementioned identification of the lengths of the links with the regulators. As a result of this restriction, the classical limit cannot be directly restored, hence the continuous limit of the theory cannot be reached.

In this paper, the flux fields modified by group averaging density distributions along links are defined. Although these fields act by adding the $\mathfrak{su}(2)$ generators at links in the same manner as the flux operators in the original LQG formulation \cite{Thiemann:2007zz}, they do not spoil the SU$(2)$ symmetry. This is possible due to their construction, which connects the separated (by the flux action) pairs of generators (in the holonomy exponential maps to the group elements). Moreover, by using the particular functionals of holonomies, the constructional modification of the flux operators does not contribute to the results of their actions. It should be emphasized that these functionals reflect the structure of the Ashtekar connections inside the graph's elementary cells \cite{Bilski:2020poi}. Moreover, they are the only contributions of the links-defined holonomies that are present in the lattice-smeared Hamiltonian constraint \cite{Bilski:2020poi,Bilski:2021_RCT_II}. Furthermore, the preserved gauge invariance should not be surprising, because the presence of intertwiners on the lattice is the result of solving the first-class constraint (known as the Gauss constraint) independently of the fluxes-consisting scalar constraint.

The introduced modification allows then to derive the continuous and classical limits of the lattice gravity exactly. In this case, although the lengths of the links remain finite, the regulators in the scalar constraint can be formally taken to zero.


\section{Algebra at the boundary}\label{Sec_Boundary}

\noindent
In this article, a simple toy model consisting of a single piecewise linear link
\begin{align}
\label{piecewise_linear}
\begin{split}
\!l\.\pos{v}:=&\;l\.\pos{v,v\@+\@l}:=l\.\pos{v,v\@+\@l/N}\@\@\circ\@\pos{v\@+\@l/N,v\@+\@2l/N}\!
\\
&\circ...\circ\@\pos{v\@+\@(N\@-\@1)/N,v\@+\@l}
\end{split}
\end{align}
is going to be considered. In the limit when the number of pieces $N$ goes to infinity, this structure approximates a smooth link arbitrarily well. Let the endpoints of the link be called the boundary and all the structure inside --- the interior.

By considering the limit $N\to\infty$ in \eqref{piecewise_linear}, the piecewise linear framework acquires the physical applicability, \textit{cf.} \cite{Thiemann:2007zz}. The lattice composed of the related piecewise links lattice is the quantity on which the gravitational theory is construable \cite{Bilski:2021_RCT_II}. Moreover, the diffeomorphism invariance on this structure is relatively easily implementable \cite{Zapata:1997da,Zapata:1997db}; thus, this latter issue is not going to be discussed in this analysis.

The canonical fields $A_a:={A_a^j\tau^j}$ and $E^a:={E^a_j\tau^j}$, where $\tau^j:={-\i\.\sigma^j/2}$ is the generator of the $\mathfrak{su}(2)$ algebra ($\sigma^j$ represents a Pauli matrix), are regulated differently. Let ${h_{l\pos{v}\@}^{-1}}:={h_{l\pos{v\!\.,v+l}\@}^{-1}}:=\prod_{j=1}^3\@\big(h_{l\pos{v}\@}^{\@(j\!\.)}\big)^{-1}$ denote the SU$(2)$ holonomy of the $\mathfrak{su}(2)$-valued connection $A_a$, where the indices in brackets are not implicitly summed. The reciprocal factors are defined by
\begin{align}
\label{holonomy_coefficient}
h_{l\pos{v}}^{(j)}\@
:=\mathcal{P}\exp\!\bigg(\!\int_{\@l\pos{v}}\!\!\!\!\!\!ds\,\dot{\ell}^a\pos{v}\@(s)\.A_a^{(j)}\@\big(l(s)\big)\tau^{(j)}\!\bigg),
\end{align}
where ${l\.\pos{v}}:={\mathbb{L}_0\varepsilon\.\pos{v}}$ is the length of the (piecewise linear) path, $\mathbb{L}_0$ represents a fiducial length scale, and ${\varepsilon\.\pos{v}}$ denotes the path-related dimensionless parameter.

The holonomy expansion around the parameter ${\varepsilon\.\pos{v}}$ can be expressed as a polynomial of the connection,
\begin{align}
\begin{split}
\label{holonomy_polynomial}
h_{l\pos{v}}^{\pm1}=&\;
\mathds{1}\pm\frac{1}{2}l\.\pos{v}\big(A_l\pos{v\@+\@l}-A_l\pos{v}\big)
\\
&+\frac{1}{2}l^2\pos{v}A_l^2\pos{v}+\mathcal{O}(\varepsilon^3)\,.
\end{split}
\end{align}
This formula is obtained by applying the derivative expansion ${\big(A_l\pos{v\@+\@l}-A_l\pos{v}\big)/l\.\pos{v}}={\partial_lA_l\pos{v}}+\mathcal{O}(\varepsilon)$ and leads to the following relation between the connections and holonomies,
\begin{align}
\label{holonomy_canonical}
h^{\mathstrut}_{l\pos{v}}\@-h_{l\pos{v}}^{\mathstrut-1}=l\.\pos{v}\big(A_l\pos{v\@+\@l}-A_l\pos{v}\big)
+\mathcal{O}(\varepsilon^3)\,.
\end{align}
This link-smeared representation of the connections located at nodes supports the construction of the lattice gravity in \cite{Bilski:2021_RCT_II,Bilski:2020poi}.

The lattice quantity that represents the densitized dreibein degree of freedom is the flux through the surface orthogonal to the spatial direction of this densitized dreibein. Let $S^{l\.\pos{v}}$ denotes the surface of area ${l^2_{\!\perp}\pos{v}}:=\epsilon^{l\.l^{\phantom{\prime}}_{\!\perp}l^{\prime}_{\!\perp}\.}\mathbb{L}_0^2\.{\varepsilon^{\phantom{\prime}}_{\!\perp}\pos{v}}\.{\varepsilon^{\prime}_{\!\perp}\pos{v}}$ normal to the link ${l\.\pos{v}}$ at the node $v$. The related flux is defined as
\begin{align}
\label{flux_coefficient}
\begin{split}
f_i\@\big(\@S^{l\.\pos{v}}\@\big)
:=&\int_{\!S^{l\pos{v}}}\!\!\!\!\!\!\!\!\enn_aE^a_i
:=\!\int_{\!S^{l\pos{v}}}\!\!\!\!\!\!\!dy\.dz\.\epsilon_{abc}E^a_i\partial_yx^b\partial_zx^c
\\
\approx&\;\varepsilon^{\phantom{\prime}}_{\!\perp}\pos{v}\.\varepsilon^{\prime}_{\!\perp}\pos{v}\mathbb{L}_0^2E^{l}_i\pos{v}\,,
\end{split}
\end{align}
where $\enn_a$ projects $E^a_i$ into the directions normal to the surface at each point. The approximation becomes the equality in the flat limit.

Next, the smearing of the densitized dreibein along the link ${l\.\pos{v}}$ is defined. The continuous probability distribution of its homogeneous linear density ${\mathcal{E}_i(l\.\pos{v})}$ reads
\begin{align}
\label{average_coefficient}
\begin{split}
\bar{\mathtt{E}}^{l\@}\big(R\@\,\pos{v}\big):=&\int_{\@l\.\pos{v}}\!\!\!\!\!\!\mathcal{E}_i(l\.\pos{v})
\\
=&\;\frac{1}{\mathbb{L}_0\.\varepsilon\.\pos{v}\!}\int\!\!dl\.\pos{v}\.E_i(l)
\approx\,\stackrel{\scriptscriptstyle\textsc{m\!\.e\!\.a\!\.n}\!}{E}{\!\!}_i^{\,l\!}\big(l\.\pos{v}\big)\,,
\end{split}
\end{align}
where ${\bar{\mathtt{E}}^{l\@}\big(R\@\,\pos{v}\big)}:=\bar{\mathtt{E}}^{l\@}\big({l\@\,\pos{v}}\!\.\wedge {S^{l\.\pos{v}}}\big)=\bar{\mathtt{E}}^{l\@}\big({l\@\,\pos{v}}\!\.\wedge{l^{\phantom{\prime}}_{\!\perp\@}\pos{v}}\!\.\wedge {l^{\prime}_{\!\perp\@}\pos{v}}\big)$. The last relation in \eqref{average_coefficient} is exact regarding the linear path, where
\begin{align}
\label{average_mean}
\stackrel{\scriptscriptstyle\textsc{m\!\.e\!\.a\!\.n}\!}{E}{\!\!}_i^{\,l\!}\big(l\.\pos{v}\big):=\frac{1}{2}\big(E^l_i\pos{v}+E^l_i\pos{v\@+\@l}\big)\,.
\end{align}
This quantity, which replaces the probability distribution in the linear path approximation, equals to the densitized dreibein that is averaged at the boundary.

%
%
\begin{figure}[h]
\vspace{-5pt}%
\begin{center}
\begin{tikzpicture}[scale=0.8]
\node at (-2.5,2.0) {$\displaystyle\frac{1}{2}$};
\draw [draw=lightgray, fill=gray, opacity=0.56] (-2.1,1.175) -- (-2.2,2.86) -- (-1.5,3.0) -- (-1.5,1.1) -- cycle; 
\draw[cb25]plot[smooth, tension=0.5] coordinates 
{(-1.5,2.0) (-1,2.025) (-0.5,2.1) (0,2.2) (0.5,2.275) (1,2.325) (1.5,2.3)};
\draw [draw=lightgray, fill=gray, opacity=0.56] (-1.5,1.1) -- (-1.5,3.0) -- (-0.7,3.16) -- (-0.7,1.0) -- cycle;
\node at (2.4,2.0) {$\displaystyle+\ \ \frac{1}{2}$};
\draw [draw=lightgray, fill=gray, opacity=0.56] (5.4,1.475) -- (5.3,3.16) -- (6.0,3.3) -- (6.0,1.4) -- cycle; 
\draw[cb25]plot[smooth, tension=0.5] coordinates 
{(3.0,2.0) (3.5,2.025) (4.0,2.1) (4.5,2.2) (5.0,2.275) (5.5,2.325) (6.0,2.3)};
\draw [draw=lightgray, fill=gray, opacity=0.56] (6.0,1.4) -- (6.0,3.3) -- (6.8,3.46) -- (6.8,1.3) -- cycle;
\end{tikzpicture}
\end{center}
\vspace{-10pt}%
\caption{Flux average at the boundary}
\label{fig_boundary}
\end{figure}
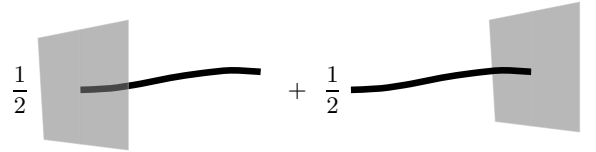
%
%
Then, the Poissson algebra regarding the canonical ADM fields \cite{Arnowitt:1960es} $q_{ab}$ and $p^{ab}$ can be derived. Considering the lattice variables intersecting at the boundary (see FIG.~\ref{fig_boundary}), one obtains
\begin{align}
\label{Poisson_average}
\Big\{h_{l\pos{v}},\!\stackrel{\scriptscriptstyle\textsc{m\!\.e\!\.a\!\.n}\!}{f}{\!\!\!\!}_i\big(\@S^{l\.\pos{v}}\@\big)\@\Big\}
=\frac{1}{2}\big(\tau^ih_{l\pos{v}}\@+h_{l\pos{v}}\tau^i\big)\,.
\end{align}
The $\mathfrak{su}(2)$ generators appeared at the nodes, \textit{i.e.} at the points, where the intertwiners implement group averaging of generators transformations. Therefore, the link (with the holonomy) remained SU$(2)$-invariant.


\section{Gauge symmetry along the link}\label{Sec_Broken_link}

\noindent
The analysis of the algebra in the interior requires more precision. Considering the flux of the ${\bar{\mathtt{E}}^{l\@}\big(R\@\,\pos{v}\big)}$ probability distribution along ${l\.\pos{v}}$, the Poisson brackets (see FIG.~\ref{fig_interior}) read
\begin{align}
\label{Poisson_distribution}
\begin{split}
\Big\{h_{l\pos{v}},\bar{f}_i^{\,\@l\!}\big(R\@\,\pos{v}\big)\Big\}
&=\lim_{N\to\infty}\!\frac{1}{2N}
\Bigg(\!
\tau^ih_{l\pos{v}}\@+h_{l\pos{v}}\tau^i
\\
+\,2&\!\sum_{n=1}^{N-1}\!h_{l\pos{v,v+n\.l/N}}\tau^ih_{l\pos{v+n\.l/N,v+l}}
\!\Bigg).
\end{split}
\end{align}
Here, the trapezoidal rule was applied to the integration along the piecewise linear path in \eqref{piecewise_linear}. Each of the elements in the sum in the lower line of \eqref{Poisson_distribution} violates the gauge symmetry.

One can demonstrate this problem explicitly. Let a point $w\in{l\.\pos{v}}$ in the link's interior be selected, hence $v\neq w\neq v+l$. Next, it is worth to define the following auxiliary group elements,
\begin{align}
{}^{^{_+}\!\!}h:=gh=hg\neq h\,,
\qquad
{}^{^{_-}\!\!}h:=g^{-1}h=hg^{-1}\neq h\,.
\end{align}
The SU$(2)$-invariant holonomy satisfies
\begin{align}
\begin{split}
gh_{l\pos{v,v+l}}g^{-1}\@=&\,{}^{^{_+}\!}{}^{^{_-}\!\!}h_{l\pos{v,v+l}}\@=\!{}^{^{_-}\!}{}^{^{_+}\!\!}h_{l\pos{v,v+l}}\@=h_{l\pos{v,v+l}}
\\
=&\;h_{l\pos{v,w}}h_{l\pos{w,v+l}}\@=\!{}^{^{_+}\!\!}h_{l\pos{v,w}}\!{}^{^{_-}\!\!}h_{l\pos{w,v+l}}\.,
\end{split}
\end{align}
for any pair $g,g^{-1}\in\text{SU}(2)$. Then, for the $\mathfrak{su}(2)$-modified interior elements in the sum in the lower line of \eqref{Poisson_distribution}, one finds
\begin{align}
\begin{split}
g\.h_{l\pos{v,w}}\tau^ih_{l\pos{w,v+l}}g^{-1}=&\;{}^{^{_+}\!\!}h_{l\pos{v,w}}\tau^ih_{l\pos{w,v+l}}^{\@\@^{_-}}
\\
\neq&\;h_{l\pos{v,w}}\tau^ih_{l\pos{w,v+l}}\,.
\end{split}
\end{align}

By adding an intertwiner at W would restore the gauge symmetry of the system; this object is a group averaging projector. It, however, becomes positioned at each node after solving one of the first-class constraints, called the Gauss constraint. The action of holonomies and fluxes is given by another one, which has to be implemented only after the Gauss constraint. This restriction occurs because the linear combination of the latter is included in the former Hamiltonian constraint. Therefore, this combination vanishes identically only after the Gauss constraint is solved. This procedure determines the constraints implementation order by the LQG technique \cite{Thiemann:1996aw,Thiemann:2007zz}. As a result, holonomies and fluxes act on the gauge-invariant lattice. Thus the action of the flux probability distribution ${{\bar{f}_i^{\,\@l\!}\big(R\@\,\pos{v}\big)}}$ is ill-defined (in the interior), in the sense that it breaks the gauge symmetry.

To restore the symmetry, the following modification of the densitized dreiben probability distribution in \eqref{average_coefficient} is introduced,
\begin{align}
\label{distribution_E}
\begin{split}
\bar{\mathtt{G}}^{l\@}\big(R\@\,\pos{v}\big)\!\.:=
&\;\frac{1}{\mathbb{L}_0\.\varepsilon\.\pos{v}\!}
\int_{\@v}^{v\@+\@l}\!\!\!\!\!\!\!\!ds\.\dot{l}(\!\.s\!\.)\.g\big(\!\.l(\!\.s\!\.)\!\.\big)E_i\@\big(\!\.l(\!\.s\!\.)\!\.\big)g^{-1}\@\big(\!\.l(\!\.s\!\.)\!\.\big)\!
\\
\approx&\,\stackrel{\scriptscriptstyle\textsc{m\!\.e\!\.a\!\.n}\!}{E}{\!\!}_i^{\,l\!}\big(l\.\pos{v}\big)\,.
\end{split}
\end{align}
The last approximation remains correct, because ${\stackrel{\scriptscriptstyle\textsc{m\!\.e\!\.a\!\.n}\!}{E}{\!\!}_i^{\,l\!}\big(l\.\pos{v}\big)}$ acts only at the nodes, \textit{i.e.} at the points, where the invariance is controlled by the intertwiners. The latter objects are adjusted for any SU$(2)$ elements at the intersecting links; hence also for the elements modified by the $g$ or $g^{-1}$ multiplication.
\begin{widetext}
%
%
\begin{center}
\begin{figure}[h]
\vspace{-10pt}%
\begin{tikzpicture}[scale=0.8]
\draw [draw=lightgray, fill=gray, opacity=0.56] (-2.1,1.175) -- (-2.2,2.86) -- (-1.5,3.0) -- (-1.5,1.1) -- cycle; 
\draw[cb25]plot[smooth, tension=0.5] coordinates 
{(-1.5,2.0) (-1,2.025) (-0.5,2.1) (0,2.2) (0.5,2.275) (1,2.325) (1.5,2.3)};
\draw[cb15,->] (-0.6,1.8) -- (-0.2,1.8);
\draw [draw=lightgray, fill=gray, opacity=0.56] (-1.5,1.1) -- (-1.5,3.0) -- (-0.7,3.16) -- (-0.7,1.0) -- cycle;
\draw [draw=lightgray, fill=gray, opacity=0.56] (2.9,1.275) -- (2.8,2.96) -- (3.5,3.1) -- (3.5,1.2) -- cycle; 
\draw[cb25]plot[smooth, tension=0.5] coordinates 
{(2.5,2.0) (3.0,2.025) (3.5,2.1) (4.0,2.2) (4.5,2.275) (5.0,2.325) (5.5,2.3)};
\draw[cb15,->] (4.4,1.9) -- (4.8,1.9);
\draw [draw=lightgray, fill=gray, opacity=0.56] (3.5,1.2) -- (3.5,3.1) -- (4.3,3.26) -- (4.3,1.1) -- cycle;
\draw [draw=lightgray, fill=gray, opacity=0.56] (7.9,1.45) -- (7.8,3.135) -- (8.5,3.275) -- (8.5,1.375) -- cycle; 
\draw[cb25]plot[smooth, tension=0.5] coordinates 
{(6.5,2.0) (7.0,2.025) (7.5,2.1) (8.0,2.2) (8.5,2.275) (9.0,2.325) (9.5,2.3)};
\draw[cb15,->] (9.4,2.075) -- (9.8,2.075);
\draw [draw=lightgray, fill=gray, opacity=0.56] (8.5,1.375) -- (8.5,3.275) -- (9.3,3.435) -- (9.3,1.275) -- cycle;
\draw [draw=lightgray, fill=gray, opacity=0.56] (12.9,1.475) -- (12.8,3.16) -- (13.5,3.3) -- (13.5,1.4) -- cycle; 
\draw[cb25]plot[smooth, tension=0.5] coordinates 
{(10.5,2.0) (11.0,2.025) (11.5,2.1) (12.0,2.2) (12.5,2.275) (13.0,2.325) (13.5,2.3)};
\draw [draw=lightgray, fill=gray, opacity=0.56] (13.5,1.4) -- (13.5,3.3) -- (14.3,3.46) -- (14.3,1.3) -- cycle;
\end{tikzpicture}
\caption{Flux probability distribution along the link}
\label{fig_interior}
\vspace{-20pt}%
\end{figure}
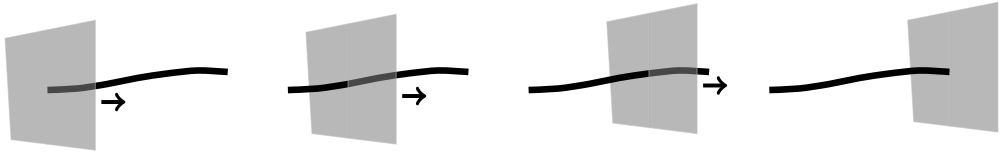
\end{center}
%
%
\end{widetext}

Analogously, one defines the gauge-invariant flux probability distribution
\begin{align}
\label{distribution_f}
\bar{g}_i^{\,\@l\!}\@\big(R\@\,\pos{v}\big)\!\.:=
\frac{1}{\mathbb{L}_0\,\@\varepsilon\,\@\pos{v}\!}\int_{\@v}^{v\@+\@l}\!\!\!\!\!\!\!\!ds\.\dot{l}(\!\.s\!\.)\.g^{-1}\@\big(\!\.l(\!\.s\!\.)\!\.\big)
\!\!\int_{\!S^{l(\!\.s\!\.)}}\!\!\!\!\!\!\!\!\!\enn_aE^a_i\@\big(\!\.l(\!\.s\!\.)\!\.\big)g\big(\!\.l(\!\.s\!\.)\!\.\big).
\end{align}
This leads to the following improvement of the algebra in \eqref{Poisson_distribution} (see also FIG.~\ref{fig_interior}),
\begin{align}
\label{distribution_algebra}
\begin{split}
\Big\{h_{l\pos{v}},\bar{g}_i^{\,\@l\!}\big(R\@\,\pos{v}\big)\Big\}
&=\lim_{N\to\infty}\!\frac{1}{2N}
\Bigg(\!
\tau^ih_{l\pos{v}}\@+h_{l\pos{v}}\tau^i
\\
+\,2\!\sum_{n=1}^{N-1}&\!\prod_{j=1}^3\@h^{(j)}_{l\pos{v,v+n\.l/N}}\tau^ih^{(j)}_{l\pos{v+n\.l/N,v+l}}
\!\Bigg).
\end{split}
\end{align}
The product in the second line expresses the transfer of the $\mathfrak{su}(2)$ generators between ${l\.\pos{v,v\@+\@n\.l/N}}$ and ${l\.\pos{v\@+\@n\.l/N,v\@+\@l}}$ in the exponential maps from representations to group elements. The gauge symmetry of each element is preserved along the whole link ${l\.\pos{v}}$, except the point ${v\@+\@n\.l/N}$. By using the abstract notation, this issue can be simply expressed in the relation
\begin{align}
g\.h_{l\pos{v,w}}g^{-1}\tau^igh_{l\pos{w,v+l}}g^{-1}=\prod_{j=1}^3\!h^{(j)}_{l\pos{v,w}}\tau^ih^{(j)}_{l\pos{w,v+l}}.\!
\end{align}


\section{Algebra along the link}\label{Sec_Algebra}

\noindent
It is convenient to introduce the auxiliary quantity that represents the $l/N$-short interval
\begin{align}
\label{notation}
\Delta l\.\pos{v_n}:=l\.\pos{\bar{v}_n-l/(2N),\bar{v}_n+l/(2N)}
\end{align}
centered at $\bar{v}_n:={v+(n-1/2)l/N}$. Thus, the lower line of \eqref{distribution_algebra} takes the form
\begin{align}
\label{Poisson_lower}
2\!\sum_{n=1}^{N-1}\prod_{j=1}^3\prod_{p=1}^n\prod_{q=n+1}^N\!\!\!h_{\Delta l\.\pos{v_{\@p\@}}}\tau^ih_{\Delta l\.\pos{v_{\@q\@}}}\,.
\end{align}
In the limit $N\to\infty$, the connection in the exponent of each $h_{\Delta l\.\pos{v_{\@p\@}}}$ is arbitrarily well approximated by the constant connection ${A^j_l\pos{\bar{v}_p}}$ that is located at the interval's center. Consequently, each holonomy can be expressed by
\begin{align}
\label{holonomy_constant}
h_{\Delta l\.\pos{v_{\@p\@}}}=\cos\@\big(\alpha^{(j)}_p\big)\mathds{1}+2\sin\@\big(\alpha^{(j)}_p\big)\tau^{(j)}\,,
\end{align}
where the notation was simplified by introducing another auxiliary variable $\alpha^j_p:=\frac{1}{2}\Delta l\.\pos{v_{(p)}}A^j_l\pos{\bar{v}_{(p)}}$. As a result, the formula in \eqref{Poisson_lower} takes the form
\begin{align}
\label{Poisson_solution}
\begin{split}
&(N-1)\big(\tau^ih_{l\pos{v}}\@+h_{l\pos{v}}\tau^i\big)
\\
+\.&(N-1)\Bigg(\!
1-\cos\!\bigg(\frac{1}{2}l\.\pos{v}\@\sum_{j=1}^3\@{}^{^{_0}\!\!\!}A_l^{\@(j)\@}\big(l\.\pos{v}\big)\@\bigg)
\!\Bigg)\@\tau^i
\\
-\.&\!\sum_{n=1}^{N-1}\sin\!\Bigg(\sum_{p=1}^n\@\alpha^i_p-\!\!\!\sum_{q=n+1}^N\!\!\!\alpha^i_q\Bigg)\,,
\end{split}
\end{align}
where ${}^{^{_0}\!\!\!}A_l^{\@(j)\@}(l\.\pos{v}):=\!\sum_{p=1}^{N}A^j_l\pos{\bar{v}_p}$. For detailed derivation see \cite{Bilski:2021_RCT_II}. The last term in \eqref{Poisson_solution} vanishes in the limit $N\to\infty$ and only the first pair of elements contributes to \eqref{distribution_algebra},
\begin{align}
\begin{split}
\label{Poisson_final}
\Big\{h_{l\pos{v}},\bar{g}_i^{\,\@l\!}\big(R\@\,\pos{v}\big)\Big\}
&=\frac{1}{2}
\Bigg(\!
\tau^ih_{l\pos{v}}\@+h_{l\pos{v}}\tau^i\@
\\
+\,2&\sin^2\!\bigg(\frac{\mathbb{L}_0\.\varepsilon^l\pos{v}\!}{4}\sum_{j=1}^3\@{}^{^{_0}\!\!\!}A_l^{\@(j)\@}\big(l\.\pos{v}\big)\!\bigg)\@\tau^i
\!\Bigg).
\end{split}
\end{align}
If the $\mathcal{O}(\varepsilon^2)$ corrections are neglected, this result coincides with the gauge-invariant expression at the boundary, given in \eqref{Poisson_average}. Moreover, despite its complicated structure, this outcome preserves the symmetry even if these corrections are included.

Eventually, one should consider the symmetric distribution of the Ashtekar connections introduced in \eqref{holonomy_canonical}. This distribution coincides with the only structure of holonomies, which is present in the Poisson brackets in the Hamiltonian constraint of the lattice gravity in \cite{Bilski:2021_RCT_II,Bilski:2020poi}. In this case, in which the regularization lattice is quadrilaterally-hexahedral, the $\mathcal{O}(\varepsilon^2)$ corrections vanish and the final result becomes
\begin{align}
\begin{split}
\label{Poisson_physical}
&\,\Big\{\Big(h^{\mathstrut}_{l\pos{v}}\@-h^{\mathstrut-1}_{l\pos{v}}\Big),\bar{g}_i^{\,\@l\!}\big(R\@\,\pos{v}\big)\Big\}
\\
=&\;\frac{1}{2}
\bigg(\@
\tau^i\Big(h^{\mathstrut}_{l\pos{v}}\@-h^{\mathstrut-1}_{l\pos{v}}\Big)
+\Big(h^{\mathstrut}_{l\pos{v}}\@-h^{\mathstrut-1}_{l\pos{v}}\Big)\tau^i
\@\bigg).
\end{split}
\end{align}
It is easy to see that the same outcome is obtained concerning the connections-holonomies relation in \eqref{holonomy_canonical} and the fluxes ${\@\stackrel{\scriptscriptstyle\textsc{m\!\.e\!\.a\!\.n}\!}{f}{\!\!\!\!}_i\big(\@S^{l\.\pos{v}}\@\big)}$ acting only at the boundary.


\section{Conclusions}\label{Sec_Conclusions}

\noindent
In the analysis in this article, it was demonstrated how the specific modification of the densitized dreibein flux representation leads to its well-defined action on links' interiors. Moreover, this result coincides with the same action at nodes if one considers the representation in \eqref{holonomy_canonical}. This latter representation appears to be physically favored due to the geometrical arguments. It is the link-defined holonomy functional in the symmetry-preserving map from the set of the continuous variables forming the Hamiltonian constraint into its lattice-smeared equivalents \cite{Bilski:2020poi}. The densitized dreibein flux functional in this map is given by the probability distribution in \eqref{distribution_f}. Thus, the whole symmetry-preserving map from the continuous into the lattice variables is self-consistent \cite{Bilski:2021_RCT_II}. This article provided another argument to favor the connections-holonomies relation in \eqref{holonomy_canonical}.

Taking the limit $\varepsilon_{H}\to0$, while keeping $\varepsilon_{\Gamma}$ small but finite becomes possible. Here, $\varepsilon_{H}$ is the Hamiltonian-related and $\varepsilon_{\Gamma}$ is the lattice-related regularization parameter, respectively. However, the removal of these two regulators identification is just a formal manipulation. The spectrum of the scalar constraint operator is not zero only on the states on which it acts, thus on the existing links. Therefore, the release of the regulators identification is only a formal method to obtain, by an exact procedure, the classical Hamiltonian from the operator. One more comment is worth to be added. It is possible to define the vacuum states for the zero-valued connections $A=0$ that lead to the vanishing holonomies by using the holonomy distribution in \eqref{holonomy_canonical} (see discussions in \cite{Bilski:2021_RCT_II,Bilski:2020poi}). The related links, however, are not going to be removed. This comment clarifies the restriction of the Hamiltonian constraint operator domain to the existing links.

Finally, in the Abelian case, the result in \eqref{Poisson_physical} equals the analogous cosmologically-reduced derivation in \cite{Bilski:2021fki}. However, it is worth noting that this latter outcome is not calculated by using the expansion in \eqref{holonomy_polynomial}. It is based on the explicit treatment of a holonomy as an SU$(2)$ group element and the holonymy's exponent as the representation of this Lie group \cite{Bilski:2020xfq}.
\\



{\acknowledgments
\noindent
This work was partially supported by the National Natural Science Foundation of China grants Nos. 11675145 and 11975203.}



\end{document}